\documentclass{aa}
\usepackage{graphicx, graphics}
\usepackage{CJK}
\usepackage{bbm}
\usepackage{amsmath, amssymb, bm}
\usepackage[nodayofweek]{datetime}
\usepackage{txfonts}
\usepackage{ulem}
\usepackage{booktabs}  
\usepackage[switch]{lineno}

\usepackage{tikz}
\definecolor{lime}{HTML}{A6CE39}

\newdateformat{monthyearday}{%
  \THEYEAR\ \monthname[\THEMONTH] \twodigit{\THEDAY}}
\newdateformat{daymonthyear}{%
  \twodigit{\THEDAY}\ \monthname[\THEMONTH] \THEYEAR}

\begin{document} 

   \title{Probing the shape of the brown dwarf desert around main-sequence A-F-G-type stars using post-common-envelope WD$-$BD binaries}


   \author{Zhangliang Chen\inst{\ref{inst-sysu}, \ref{inst-csst}}
   \and 
   Yizhi Chen\inst{\ref{inst-sysu}, \ref{inst-csst}} 
   \and 
   Chen Chen\inst{\ref{inst-sysu}, \ref{inst-csst}} 
   \and 
   Hongwei Ge\inst{\ref{inst-YNO}, \ref{inst-KLS},\ref{inst-ICS}}
   \and
   Bo Ma\inst{\ref{inst-sysu}, \ref{inst-csst}}  \fnmsep\thanks{Corresponding author}
          }

   \institute{School of Physics and Astronomy, Sun Yat-sen University,
Zhuhai 519082, People's Republic of China;
              \email{\url{mabo8@mail.sysu.edu.cn}}  \label{inst-sysu} %
         \and
         CSST Science Center for the Guangdong-HongKong-Macau Great Bay Area, Sun Yat-sen University, Zhuhai 519082, People's Republic of China \label{inst-csst}
         \and
             Yunnan Observatories, Chinese Academy of Sciences, 396 YangFangWang, Guandu District, Kunming, 650216, People's Republic of China
             \email{gehw@ynao.ac.cn} \label{inst-YNO}
        \and
            Key Laboratory for Structure and Evolution of Celestial Objects, Chinese Academy of Sciences, P.O. Box 110, Kunming 650216, People's Republic of China \label{inst-KLS}
        \and
            International Centre of Supernovae, Yunnan Key Laboratory, Kunming 650216, People's Republic of China \label{inst-ICS}
             }

   \date{Received ; accepted }

  \abstract
   {
   Brown dwarfs (BDs) with masses in the range 40$-$60 $M_{\rm Jup}$ are rare around solar-type main-sequence (MS) stars, which gives rise to the brown dwarf desert (BDD). One caveat associated with previous studies of BDD is the relatively limited sample size of MS$-$BD binaries with accurately determined BD masses.}
  { We aim to produce a large sample of BD companions with precisely determined masses around MS A-F-G-type stars using observations of post-common-envelope white dwarf (WD)$-$BD binaries.
  }
  {We employed the rapid binary evolution code COMPAS to deduce the properties of MS$-$BD binary progenitors from post-common-envelope WD$-$BD binaries. 
  With this method, we increase the sample of  directly observed MS$-$BD binaries, enriching the data available for analyzing the BDD around MS A-F-G-type stars.
  }
  { Our study opens a new window for studying the shape of the BDD around A-F-G-type MS stars in the short-period regime. We find tentative evidence, albeit with a small sample size, that the ``driest'' part of the BDD around A-F-G-type stars may extend into an orbital period of several hundred days. 
  More post-common-envelope WD$-$BD binaries detected in the future will advance our understanding of the BDD around A-F-G-type stars. 
  }
  {}

   \keywords{brown dwarfs -- white dwarfs -- binaries: close -- methods: statistical}

\authorrunning{Chen et al.}
\titlerunning{Probing the brown dwarf desert using post-CE WD$-$BD binaries}

   \maketitle
%

\section{Introduction}
Brown dwarfs (BDs) are substellar objects with masses in the range 13-75~$M_{\rm Jupiter}$, filling the gap between giant planets and low-mass stars \citep{Burrows01}.
A notable deficit of BDs has been observed around main-sequence (MS) solar-type stars, which is designated the ``brown dwarf desert'' \citep[BDD;][]{Halbwach00,Marcy00}. \citet{Farihi05} suggest that white dwarf (WD) systems may also exhibit a similar scarcity of BD companions.
The formation of such a desert is believed to be linked to the different formation mechanisms of giant planets and low-mass stars, although the mass boundary between these two classes of objects remains a subject of active debate \citep{Burrows01, Grether06, Persson19, Kiefer21, Feng22, Stevenson23}.
Therefore, examining the shape of the BDD can provide information valuable for the understanding of the planetary and star formation mechanisms around MS stars \citep{Ma14, Shahaf19}. 

However, the majority of BDs analyzed in previous studies lack precise mass measurements. 
Similar to the detection of exoplanets, both the transit and radial velocity (RV) techniques have been used to detect short-period MS$-$BD binaries \citep{Grieves17, Lin23}.
Due to the low occurrence rate of BDs around MS solar-type stars, there are only about two hundred known short-period MS$-$BD binaries \citep{Stevenson23}. 
Of these binaries, only a few dozen have precise mass measurements from a combination of transit, RV, and astrometry observations \citep[see][and references therein]{Grieves21, Feng22, Stevenson23}, and the rest only have $\rm m\sin(i)$ values measured using the RV technique \citep{Kiefer21}.
To better constrain the location and shape of the BDD, a larger BD sample with precise mass measurements is essential. 

The discovery of gravitational waves (GWs) from GW170817 with Laser Interferometer Gravitational-Wave Observatory (LIGO) marked the start of GW astronomy \citep{GW170817}. Subsequently, the allocation of telescope resources has been directed toward the search for WD binaries in the Milky Way via photometry and RV methods. These binaries serve as critical targets for forthcoming space-based GW observatories, such as Laser Interferometer Space Antenna \citep[LISA,][]{Lisa}, Tianqin \citep{tianqin}, and Taiji \citep{taiji}.
For instance, \citet{Burdge20} have identified several extremely short-period WD$-$WD binaries, deemed ideal targets for these GW missions.
While WD$-$BD binaries may not exhibit the strong GW emission targeted by these next-generation observatories, discoveries of such binaries have nonetheless emerged from GW surveys \citep{Parsons17,Casewell18EPIC}. 
Despite the fact that only a dozen such systems have been detected, \citet{Zorotovic22} have demonstrated the feasibility of employing their evolutionary histories to constrain the common-envelope (CE) efficiency. 

In this study we present an additional application of these WD$-$BD binaries. 
We propose using the WD$-$BD binary sample as a means to investigate the shape of the BDD around MS A-F-G-type stars. 
Utilizing the rapid binary evolution code COMPAS, we can determine the progenitor binary system from the observed WD$-$BD binary, which is often an A-F-G-type MS$-$BD binary. 
This new MS$-$BD progenitor sample is an excellent complement to the current MS$-$BD sample used for probing the shape of the BDD around solar-type stars, as these new MS$-$BD progenitors occupy a totally different regime in the period-mass diagram. 
In Section~\ref{sec:data} we present the data used in our analysis. The method and results are presented in Section~\ref{sec:method}. We provide a brief discussion in Section~\ref{sec:disc} and summarize our results in Section~\ref{sec:conc}.

\begin{table*}[ht!]
        \centering
        \caption{ Parameters of WD$-$BD binaries from observations. }
        \label{tab:ob_WDBD}
        \begin{tabular}{lccc} 
                \hline
                System & M$_{\rm WD}$ & M$_{\rm BD}$ & Period\\
        & ($M_{\odot})$ & ($M_{\odot}$) & (min) \\
                \hline
        SDSS~J1411$+$2009 & 0.53±0.03 & 0.050±0.002 & 121.7\\
        SDSS~J1205$-$0242 & 0.39±0.02 & 0.049±0.006 & 71.2\\
		WD~1032$+$011 & 0.45±0.05 & 0.067±0.006 & 131.8\\
		ZTF~J0038$+$2030 & 0.50±0.02 & 0.0593±0.004 & 622.0\\
        WD~0137$-$349 & 0.39±0.035 & 0.053±0.006 & 115.6\\
        NLTT~5306 & 0.44±0.04 & 0.053±0.003 & 101.9\\
        SDSS~J1557$+$0916 & 0.447±0.043 & 0.063±0.002 & 136.4\\
        EPIC~212235321 & 0.47±0.01 & $0.055^{+0.007}_{-0.010}$ & 68.2\\
                \hline
        \end{tabular}
\end{table*}

\section{Data and observation} \label{sec:data}

To demonstrate the feasibility of employing the WD$-$BD binary distribution to probe the shape of the BDD around MS stars, we gathered data of eight known close WD$-$BD binaries with precise mass measurements from the literature. 
The parameters of these WD$-$BD binaries are summarized in Table~\ref{tab:ob_WDBD}. 
The discovery and orbital parameters for each of these systems have been reviewed by \citet{Zorotovic22}.
We demonstrate in Section~\ref{sec:method} that all of these binaries are post-CE binaries. 
Most of the selected binaries are eclipsing systems detected from ground-based photometry survey. Because more accurate parameters, especially the masses of BDs, can usually be derived from spectral RV measurements when comparing to non-eclipsing systems.
The WDs in these systems typically possess a mass of approximately 0.5$M_{\odot}$, suggesting an F-type progenitor star with a mass of approximately 1.5$M_{\odot}$. 

\begin{figure}[ht!]
\centering
\includegraphics[scale=0.3]{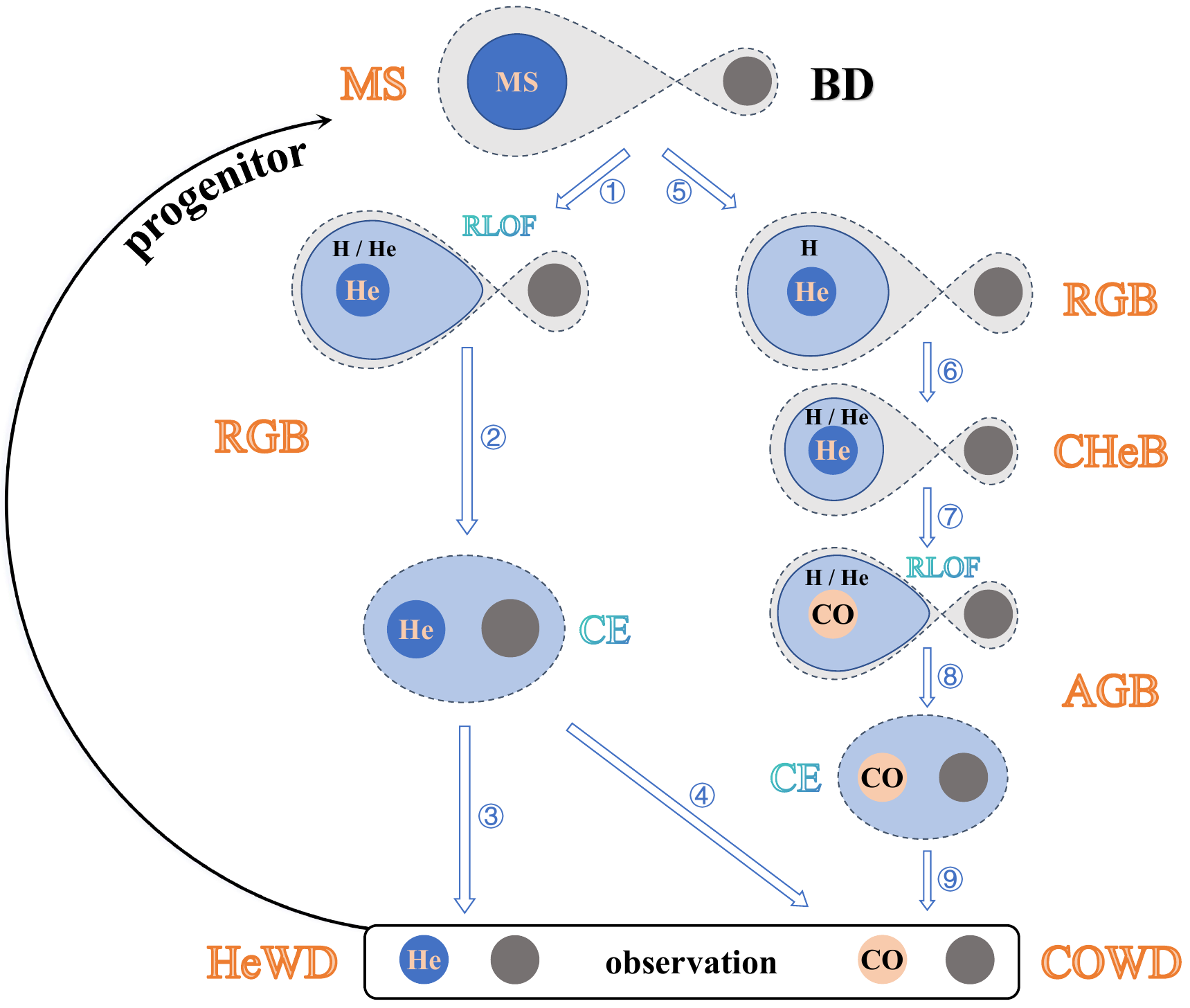}
\caption{
Formation scenarios of WD$-$BD binaries from MS$-$BD binaries. We find two main formation channels from our population synthesis simulation,  
Channel A ($1\rightarrow2\rightarrow3$ or $1\rightarrow2\rightarrow4$) and Channel B ($5\rightarrow6\rightarrow7\rightarrow8\rightarrow9$).
Using population synthesis calculations, we can infer the properties of the progenitor MS$-$BD binaries from the properties of the end WD$-$BD binaries.
}
\label{fig:evo}
\end{figure}

\section{Method and result} \label{sec:method}
\subsection{Population synthesis} \label{sec:method1}

We used a population synthesis method to derive the properties of progenitor systems for the known close WD$-$BD binaries. 
We used the code COMPAS, designed for fast population synthesis \citep{Riley22teamCOMPAS}, to calculate the evolution of the progenitor binaries from MS stage to the end WD$-$BD binaries. 
As shown in Table~\ref{tab:ob_WDBD}, most of the WDs observed in close binaries are in the mass range of $\sim$0.4 to 0.55$M_{\odot}$, and the BDs are in the mass range of 0.04 to 0.07$M_{\odot}$, with orbital periods in the range of 60 to 700 minutes. 
The formation of these close WD$-$BD binary usually involves the CE events, which can reduce the orbital periods significantly.

To cover the parameter space for all the WD$-$BD binaries in our sample, we set the initial parameter range for the progenitor binaries as follows:
the primary stellar mass is in the range of 0.8~$M_{\odot}$ to 4.0~$M_{\odot}$, 
the secondary stellar mass is in the range of 0.04~$M_{\odot}$ to 0.07~$M_{\odot}$,
the semimajor axis is in the range of 0.1~AU to 3.0~AU,
and the CE parameter, $\alpha_{\rm CE}$, is in the range of 0.05 to 1.0.
Besides, in all samples, the metallicity is set to the solar metallicity of Z=0.02, and the eccentricity is set to 0.

As the primary star expands in the red giant branch (RGB) or the asymptotic giant branch (AGB) stage, mass transfer can happen due to Roche lobe overflow (RLOF).
Following the formula in \citet{Eggleton83}, the ratio of the equivalent Roche lobe (RL) radius, $R_{\rm RL}$, to the semimajor axis, $a$, can be calculated as
\begin{equation}\label{eq:RL}
    r_{\rm RL} = \frac{R_{\rm RL}}{a} = \frac{0.49q^{2/3}}{0.6q^{2/3}+ \rm ln(1+\it q^{\rm 1/3})}.
\end{equation}
Sometimes, the mass transfer is not stable. 
The unstable system can transfer mass in a dynamical timescale and enter a classical CE phase. 
Recently, \citet{Ge20b} found an initial thermal timescale mass transfer for late RGB or AGB stars may also contribute essentially to CE evolution. The thermal timescale can be less than a hundred years for these stars, and a donor star can overfill its outer Lagrangian radius significantly. \citet{Henneco23} further found the onset of runaway mass transfer and L2 overflow often occur quasi-simultaneously in Case-Bl (convective envelope) and Case-C binaries. However, some Case-Bl and Case-C systems do not experience runaway mass transfer but do have primary stars that extend far beyond the L2-lobe, so they expect the onset of a classical CE phase.
In our simulation, we used the mass$-$radius exponents $\zeta \equiv d\ln R / d\ln M $ model from \citet{Soberman97} to determine the mass transfer phase, which compares the response of the stellar radius $\zeta_{*} \equiv d\ln R_{*} / d\ln M $ and the response of the stellar RL radius $\zeta_{\rm RL} \equiv d\ln R_{\rm RL} / d\ln M $ during mass transfer process.

Once an unstable mass transfer happens, the MS$-$BD binary will soon evolve into a CE phase \citep{Livio88,XuLi10}.
Then, the envelope might be ejected through interaction with the binary, which leads to an orbital decay of the binary \citep{Taam00}. 
Following the CE phase, the system emerges as a WD$-$BD binary.
There are two popular methods for parameterizing the CE ejection process: the $\alpha$ formalism \citep{Webbink84,Livio88,deKool90,Dewi00} and the $\gamma$ formalism \citep{Nelemans05}.
In the $\alpha$ formalism, whether the CE can be ejected depends on the relationship between the envelope binding energy, $E_{\rm bind}$, and the orbital energy, $E_{\rm orb}$. 
If the CE can be ejected successfully, the amount of orbital energy released should be equal to the binding energy of the envelope based on the conservation of energy:
\begin{equation}\label{eq:alpha}
    E_{\rm bind} = \alpha_{\rm CE}\Delta E_{\rm orb} = \alpha_{\rm CE}(E_{\rm orb,\ f}-E_{\rm orb,\ i}),
\end{equation}
where $\alpha_{\rm CE}$ is the energy conversion efficiency parameter between $E_{\rm bind}$ and $E_{\rm orb}$. Following the model in \citet{deKool90}, we used the structure parameter, $\lambda$, to evaluate the binding energy:
\begin{equation}\label{eq:bind}
    E_{\rm bind}=\frac{GM_{\rm 1}M_{\rm 1,e}}{\lambda R_{\rm 1}}=\frac{GM_{\rm 1}M_{\rm 1,e}}{\lambda a_{\rm i} r_{\rm L}},
\end{equation}
where $M_1$ and $M_{\rm 1,e}$ are the total mass and envelope mass of the primary star, $R_1$ is the radius of the primary star, $r_{\rm L}=R_{\rm L}/a_{\rm i}$ is the ratio of the RL radius and the orbital separation at the onset of CE, and $a_{\rm i} r_{\rm L}$ is normally taken as the stellar radius once a star fills its RL and starts mass transfer.
In our population synthesis code, we use the mass-interpolation values of $\lambda=\lambda_{\rm b}$ using the prescription from \citet{XuLi10}, which includes the contribution of stellar internal thermal energy. 
Additionally, the orbital energy of BD can be calculated as\begin{equation}\label{eq:orb}
    E_{\rm orb,\ i/f}=\frac{GM_{\rm 1/1,c}M_{\rm BD}}{2a_{\rm i/f}},
\end{equation}
where ${a_{\rm i}}$ and ${a_{\rm f}}$ refer to the initial and final orbital separation of the CE phase, $M_{\rm 1}$ is the mass of the primary star immediately before the CE, and $M_{\rm 1,c}=M_1-M_{\rm 1,e}$ is the core mass (also the post-CE mass) of the primary star.
By combining the Eqs.~\ref{eq:alpha}, \ref{eq:bind}, and \ref{eq:orb}, we have
\begin{equation} \label{eq:conse1}
    \frac{GM_{\rm 1}M_{\rm 1,e}}{\lambda R_{\rm 1}}=
    \alpha_{\rm CE}(\frac{GM_{\rm 1,c}M_{\rm BD}}{2a_{\rm f}}-\frac{GM_1M_{\rm BD}}{2a_{\rm i}}).
\end{equation}
From Equation~\ref{eq:conse1}, the post-CE orbital separation can be related to the pre-CE orbital separation as
\begin{equation}
    \frac{a_{\rm f}}{a_{\rm i}}=\frac{M_{\rm 1,c}M_{\rm BD}}{M_1}\frac{1}{M_{\rm BD}+2M_{\rm 1,e}/(\alpha_{\rm CE}\lambda r_{\rm RL})}.
\end{equation}

Wind loss prescription from \citet{Belczynski10} is used as default in COMPAS, which is based on Monte Carlo radiative transfer simulations of \citet{Vink01}. However, wind accretion is not currently included in COMPAS. Assuming the commonly used Bondi-Hoyle-Lyttleton wind accretion scheme \citep{Bondi44}, we used the equation from \citet{Hurley02} to estimate the wind accretion rate. We find that about $10^{-5}$ of the wind mass loss of the primary star can be accreted by the BD companion, which can be negligible for the purpose of this study.

The corresponding parameters of the WD$-$BD binaries generated using the above initial parameter range are as follows: 
the WD mass is in the range of 0.3~$M_{\odot}$ to 0.9~$M_{\odot}$,
the BD mass is in the range of 0.04~$M_{\odot}$ to 0.07~$M_{\odot}$,
and the orbital period is in the range of $10$~mins to $10^6$~mins.
After examine the results of our population synthesis simulation and the results of \citet{Zorotovic22}, we find two main channels for the formation of close WD$-$BD systems. We summarize these two formation channels in Figure~\ref{fig:evo}.
In channel A, the CE formed at the RGB stage has been fully ejected with a helium core left. If the core is massive enough, it might ignite the helium core and become a hot subdwarf B star \citep{han02,Arancibia24}. Therefore, the primary star can eventually evolve to a helium-core WD (HeWD) or a carbon$-$oxygen-core WD (COWD) in this case.  
In channel B, the primary star has not fully filled the RL radius during the RGB stage, and the binary enters the CE phase at the AGB stage after the core$-$helium-burning (CHeB) stage. The primary star then finally becomes a COWD after the ejection of CE. 

\begin{figure}[h!]
\centering
\includegraphics[scale=0.38]{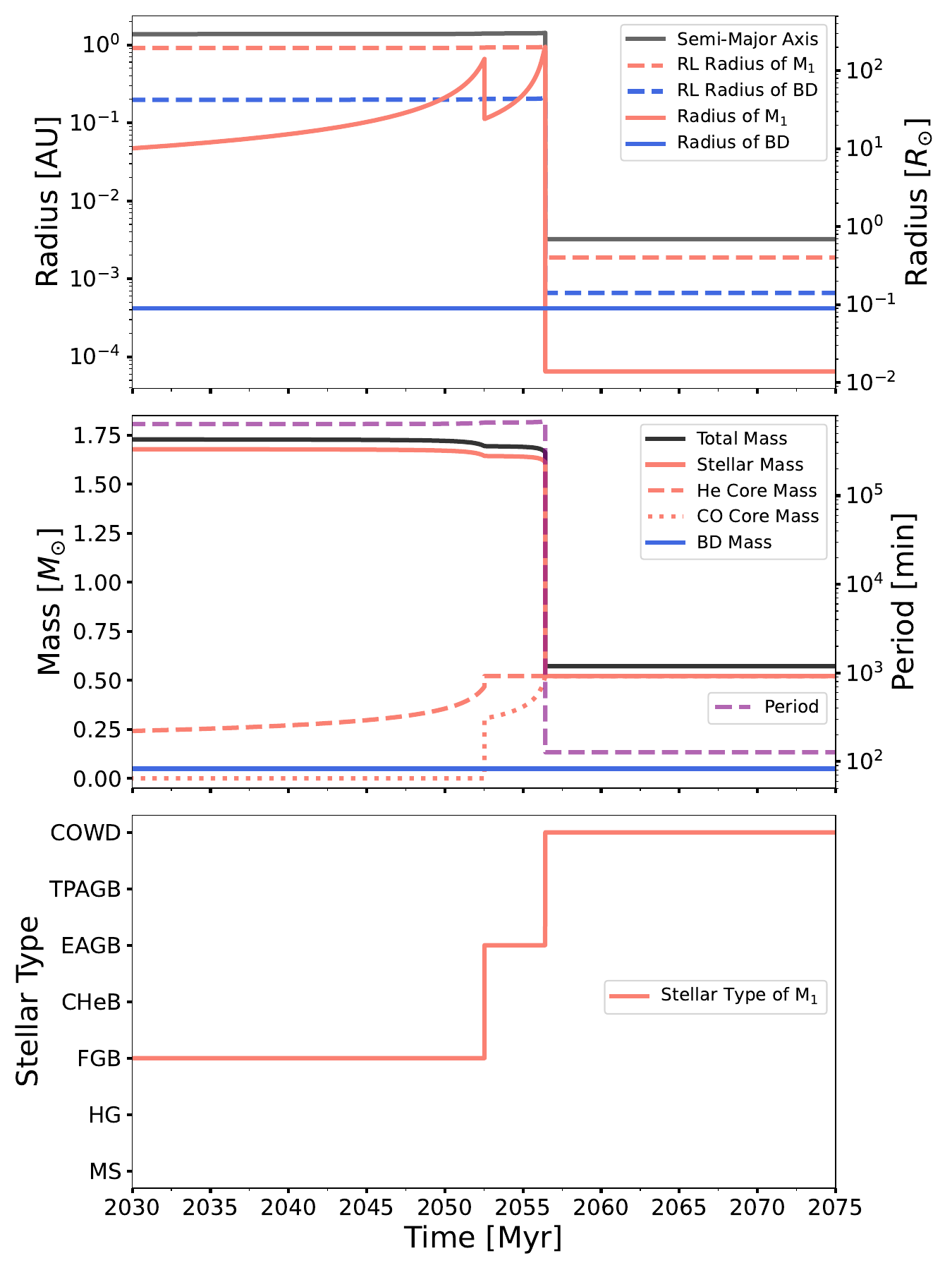}
\caption{Example evolutionary track for the formation of the WD$-$BD binary SDSS~J1411+2009. The binary consists of two stars orbiting each other, with initial masses of 1.68$M_\odot$ and 0.050$M_\odot$ and an initial orbital period of 445~days. The evolution of the stellar radius and orbital separation is shown in the top panel, the evolution of the stellar mass, core mass, and orbital period is shown in the middle panel, and the stellar type evolution of the primary star is shown in the bottom panel.}
\label{fig:sdss1411}
\end{figure}

We show an example evolutionary track for the formation of SDSS J1411+2009 in Figure~\ref{fig:sdss1411}. In this plot, we do not show the MS stage of the evolution and instead focus on the binary evolution after the primary star entering the RGB stage. 
This WD$-$BD system is more likely to be formed through channel B shown in Figure~\ref{fig:evo}. 
The initial binary separation is far enough that the BD can escape from being engulfed during the RGB stage of the primary star. 
The BD and the primary star then enter the CE phase during the AGB stage, and finally become a short-period COWD$-$BD binary after ejecting the CE.

\subsection{Binary progenitors}\label{sec:BP}
We used the stellar masses and orbital periods summarized in Table~\ref{tab:ob_WDBD} as input parameters to search for their matching progenitor systems from our population synthesis results.  
To put more realistic constraints on our reconstruction results, we followed \citet{Zorotovic22} and utilized the cooling time of WDs and the minimum total age of the WD-BD systems. 
We used the cooling time to evaluate the orbital period immediately after the CE phase, assuming that their orbital decay was driven solely by the GW emission during the post-CE stage. The minimum total age of the system can be used to constrain the maximum initial stellar mass of the primary. 
We report the matching progenitor's parameters in Table~\ref{tab:ini_WDBD}. 
We also present the orbital period immediately before the CE phase in Table~\ref{tab:ini_WDBD}, denoted as $\rm Period_{\rm pre-CE}$, to facilitate comparisons with the results of \citet{Zorotovic22}. 

\begin{figure}[h!]
\centering
\includegraphics[scale=0.35]{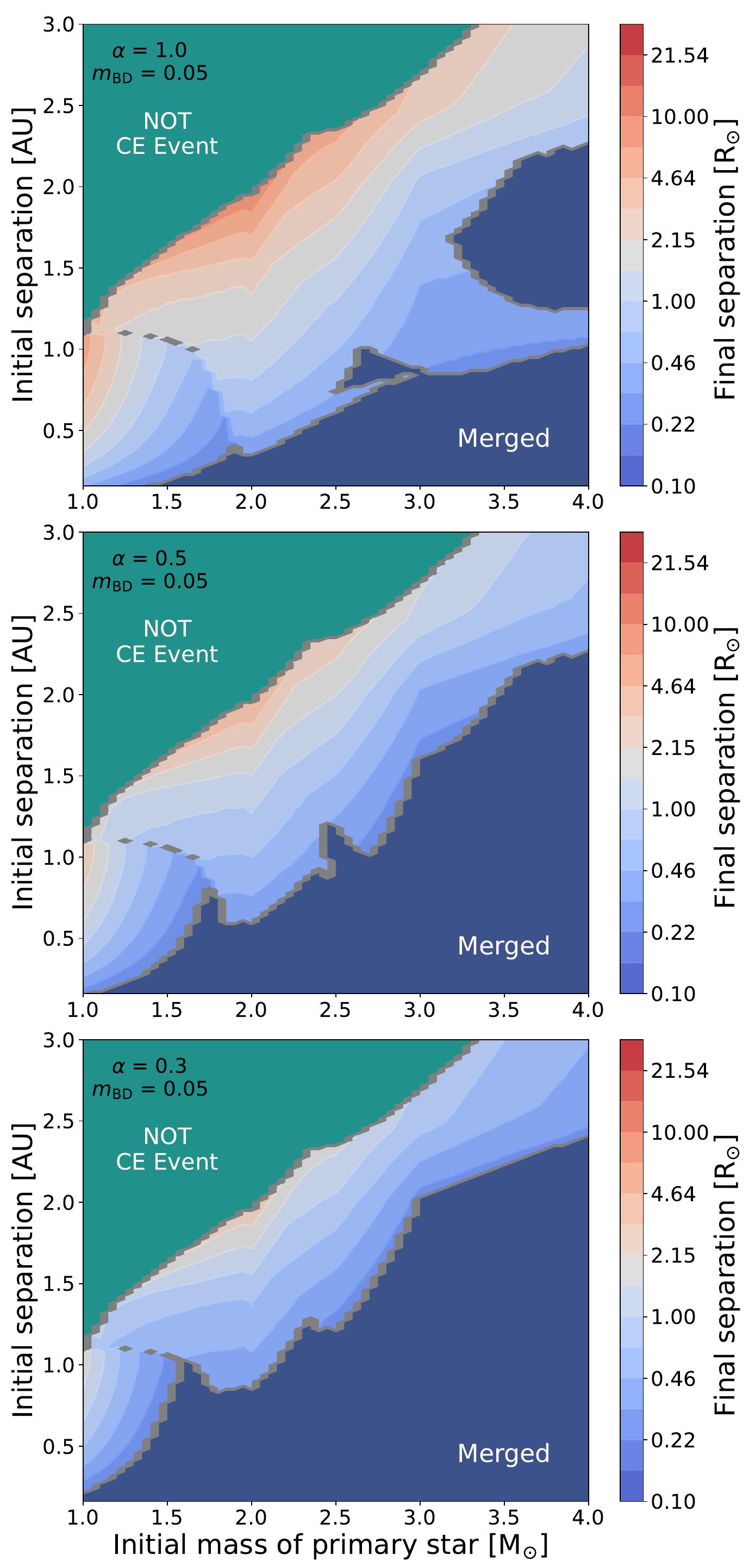}
\caption{Final separation of a WD$-$BD binary as a function of the initial mass of the primary star and the initial orbital separation, with a BD mass of 0.050~$M_{\odot}$, under the assumption that $\alpha = 0.3, 0.5, 1,$ and with an eccentricity of 0. 
The green area indicates the binaries that have never undergone a CE phase and will become wide binaries.
The dark blue area marks the binaries that will merge during and after
the CE phase, and the light blue and pink area denotes the binaries capable of forming close WD$-$BD binaries following a CE phase. The final WD$-$BD orbital separation, expressed in units of solar radii ($R_{\odot}$), is illustrated using the color bar.
}
\label{fig:m2_slice}
\end{figure}

The same WD$-$BD binary system can be reconstructed using different combinations of stellar mass and orbital period parameters. 
For example, with $\alpha_{\rm CE}$ varying from 0.14 to 0.90, we find the primary mass and initial orbital period of the progenitor binary of SDSS~J1205$-$0242 fall in the range of 1.00 to 1.48~$M_{\odot}$ and 72 to 178~days.
To investigate how the parameters of the progenitor binary and $\alpha_{\rm CE}$ values can impact the end WD$-$BD properties, we plot the distribution of final separation of WD$-$BD binaries as a function of the initial mass of the primary star and initial orbital separation in Figure~\ref{fig:m2_slice}. We use a BD mass of 0.05$M_{\odot}$ and $\alpha_{\rm CE}$ values of 0.3, 0.5, and 1.0 in this plot.
The green area shows the binaries that have never undergone a CE phase or even a mass transfer process, which will become wide binaries or even unbound \citep{Nordhaus13}. 
The dark blue area indicates the binaries that will merge during and after the CE phase to become single stars. 
The light blue and pink area denotes the binaries capable of forming close WD$-$BD binaries following a CE phase. 
By comparing results using different $\alpha_{\rm CE}$, we find a larger $\alpha_{\rm CE}$ can reduce the ``merged'' area and make the light blue and pink area larger because more orbital energy can be used to eject CE.
We also find that a larger initial orbital separation is required to form a close WD$-$BD system for a binary with a larger initial primary stellar mass. 
This is because a larger primary star means larger binding energy, and more orbital energy is needed to eject the CE.  
Furthermore, for close WD$-$BD systems, most show the tendency that a wider initial separation results in a wider final separation. 
For binaries with an initial primary mass smaller than $\sim$1.8~$M_{\odot}$ and an initial orbital separation smaller than $\sim$1.0~AU, they are more likely to evolve through Channel A shown in Figure~\ref{fig:evo}.
If the initial separation is larger than $\sim$1.0~AU, systems with the initial primary stellar mass from 1.0 to 4.0~$M_{\odot}$ will go through Channel B to form close COWD$-$BD binaries.

We also considered the latest mass transfer criteria proposed by \citet{Ge20a} in our binary evolution simulation, which gives the critical mass ratios for dynamically unstable mass transfer. 
Adopting Ge's criteria can make the boundary of the ``not CE event'' region shown in Figure~\ref{fig:m2_slice} rougher, 
which is because some corrections in COMPAS should be applied to Ge's model when dealing with thermally pulsing (TP)$-$AGB stars \citep[see, e.g.,][]{Ge23}. 
The corrections are because the ultrashort thermal timescale mass transfer through the outer Lagrangian point plays a more vital limit than the dynamical timescale mass transfer \citep{Ge20b,Ge23}.  
However, the new criteria make a lesser impact on the evolution of systems that have experienced CE. Thus, we chose to not adopt Ge's criteria in this study. 

\begin{table*}[ht!]
        \centering
        \caption{ Parameters of MS$-$BD binaries inferred from WD$-$BD binaries using a population synthesis technique. The initial period refers to the orbital period when the primary star starts its MS evolution. In our calculation, we have added minimum total age constraints on these systems.
 }
        \label{tab:ini_WDBD}
        \begin{tabular}{lcccc} 
                \hline
                 System & M$_{\rm MS}$ & Initial Period & Period$_{\rm pre-CE}$ & $\alpha_{CE}$ \\
                 & /$M_{\odot}$ & /day & /day\\
                \hline
	    SDSS~J1411$+$2009 & $1.02\sim1.44$ & $336\sim583$ & $387\sim978$ & $0.12\sim0.41$ \\
        SDSS~J1205$-$0242 & $1.00\sim1.48$ & $72\sim178$ & $74\sim219$ & $0.14\sim0.90$  \\
        WD~1032$+$011 & $1.00\sim1.24$ & $137\sim430$ & $150\sim721$ & $0.09\sim0.45$ \\
	    ZTF~J0038$+$2030 & $1.00\sim1.10$ & $402\sim486$ & $738\sim1004$ & $0.45\sim0.95$ \\
        WD~0137$-$349 & $1.00\sim1.56$ & $65\sim213$ & $68\sim270$ & $0.16\sim1.00$ \\
        NLTT~5306 & $1.00\sim1.22$ & $136\sim407$ & $149\sim663$ & $0.10\sim0.45$ \\
        SDSS~J1557$+$0916 & $1.00\sim1.72$ & $123\sim409$ & $127\sim667$ & $0.09\sim1.00$ \\
        EPIC~212235321 & $1.08\sim1.56$ & $264\sim389$ & $280\sim568$ & $0.09\sim0.49$ \\
    \hline
    \end{tabular}
\end{table*}

\begin{figure*}[h!]
\centering
\includegraphics[scale=0.6]{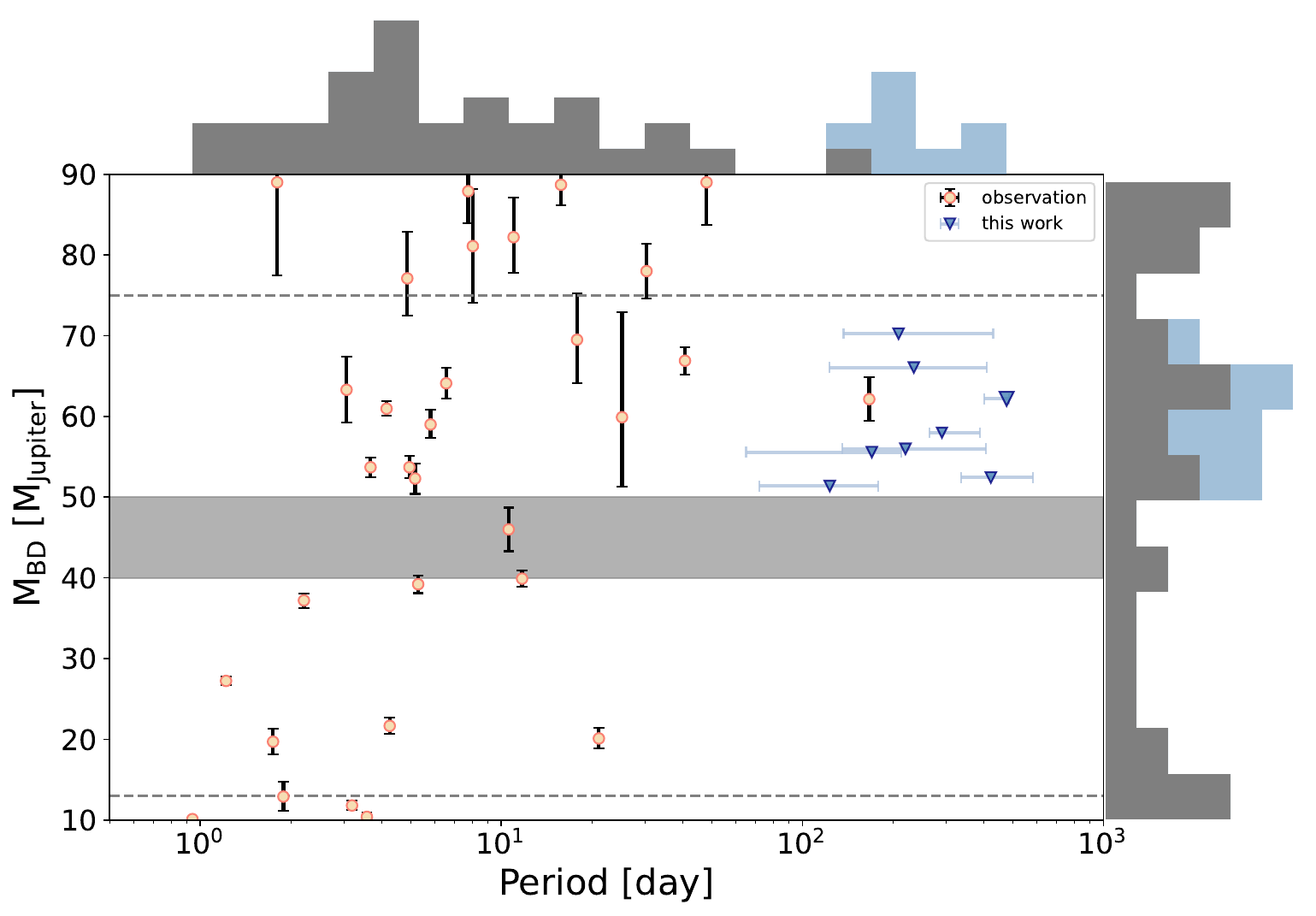}
\caption{Period-mass distribution of close MS$-$BD binaries inferred from close WD$-$BD binaries using population synthesis technique. 
We have also overplotted the known transiting A-F-G-type MS$-$BD binaries with precise BD mass measurements (excluding those only having $M\sin i$ measured) from the literature 
\citep{Pont06,Johns08,Winn08,Deleuil08,Hebrard08,Winn09,Hellier09,Triaud10,Borucki11,Anderson11,Ford11,Bouchy11a,Bouchy11b,Siverd12,Triaud13,Moutou13,Diaz13,Diaz14,Parviainen14,Ma14,Bonomo15,Csizmadia15,Nowak17,Bayliss17,Hodzic18,Zhou19,Boetticher19,Carmichael19,Persson19,Subjak20,Mireles20,Maire20,Carmichael20,Grieves21,Carmichael21,Benni21,Acton21,Stevenson23}.
Progenitor systems are marked with blue triangles based on a $\alpha_{\rm CE}$ value of 0.3 except for ZTF~J0038+2030, where a $\alpha_{\rm CE}$ value of 0.45 is used.  
The error bars illustrate the range of initial orbital periods based on all possible $\alpha_{\rm CE}$. The gray-shaded area marks the lowest density area in the BDD proposed by previous studies.}
\label{fig:general}
\end{figure*}

\section{Discussion} \label{sec:disc}
\subsection{Comparison with previous studies} \label{sec:disc1}
Parameters of progenitor binaries have been estimated in some of the WD$-$BD discovery papers \citep[see, e.g.,][]{Casewell12}.
Furthermore, \citet{Zorotovic22} have employed the WD$-$BD samples to study the CE parameter $\alpha_{\rm CE}$. They have combined the evolution of a single star using the Single Star Evolution (SSE) code from \citet{Hurley00} with the $\alpha$ formalism calculation, to reconstruct the progenitor systems of the observed WD$-$BD binaries. 
It is worth noting that the ``initial orbital period'' parameter shown in their Table~2 represents the orbital period immediately before entering the CE phase, which corresponds to the $\rm Period_{pre-CE}$ parameter in our Table~\ref{tab:ini_WDBD}. 
While in our Table~\ref{tab:ini_WDBD}, the ``Initial Period'' represents the orbital period at the beginning of the MS stage of the primary star.
For example, the initial orbital periods of the progenitor systems of SDSS~J1411$+$2009 fall in the range of 336 to 581 days. 
Since mass loss from the primary star during its giant branch stage can take away orbital angular momentum, this orbital period range will increase to 387 to 949 days before entering the CE phase according to our calculation. 
The orbital expansion caused by the mass-loss is calculated assuming an angular momentum conservation of
\begin{equation}
    \frac{\dot{a}}{a} = -\frac{\dot{M_{\rm tot}}}{M_{\rm tot}},
\end{equation}
where the $M_{\rm tot}$ represents the total mass of the system. 
Our orbital period range immediately before entering CE agrees well with the range calculated by \citet{Zorotovic22} for SDSS~J1411$+$2009, which is from 334 to 937 days.

The slight discrepancy in the orbital period range before entering CE likely arises from the use of different binding energy parameter and different formulation for the orbital energy. 
For example, \citet{Zorotovic22} used the structural parameter $\lambda$ published by \citet{Claeys14} in the last version of the Binary star evolution (BSE) code, while we used $\lambda_b$ values from \citet{XuLi10} in COMPAS. The initial orbital energy was calculated between the secondary star and the whole primary in COMPAS, while in \citet{Zorotovic22} the initial orbital energy was calculated between the secondary star and the core of the primary star.

\subsection{$\alpha$ formalism efficiency} \label{sec:disc3}
The efficiency parameter $\alpha_{\rm CE}$ in the $\alpha$ formalism has been estimated to be between $0.2$ and $0.4$, by reconstructing the evolution of a population of WD$-$WD, WD$-$MS, or WD$-$BD binaries \citep[][]{Zorotovic10,Toonen13,Camacho14,Parsons15,Hernandez21,Hernandez22,Zorotovic22, Scherbak23}.
Similar to \citet{Zorotovic22}, when not using a minimum age limit, $\alpha_{\rm CE}=1$ can explain the formation of all the systems with the exception of EPIC~212235321. But when adding a minimum age limit, we find using an $\alpha_{\rm CE}$ value of $\sim$0.2 to 0.4 can reproduce almost all of the known WD$-$BD, except for ZTF~J0038+2030. 
Given the consistency in mass and period ranges between the two studies, the different $\alpha_{\rm CE}$ ranges for ZTF~J0038+2030 likely arise from the use of different $\lambda$ values. 
Thus, the study of binding energy parameter $\lambda$ of the CE evolution is also very import in constraining the $\alpha_{\rm CE}$ values. 
With more and more WD$-$BD binaries discovered in the future, we expect to have a better understanding of the distribution of $\alpha_{\rm CE}$.



\subsection{Period-mass distribution for A-F-G-type stars} \label{sec:disc2}
Previous studies have used the period-mass distribution to study the shape of the BDD \citep{Shahaf19}. 
In Figure~\ref{fig:general} we plot the period-mass distribution of all the progenitor systems for the WD$-$BD systems derived in Sect.~\ref{sec:BP}. 
We have also over-plotted all known transiting or eclipsing A-F-G-type MS$-$BD systems with precise BD mass measurements in Figure~\ref{fig:general}. 
Based on previous studies, the efficiency parameter, $\alpha_{\rm CE}$, has been estimated to be in the range of $0.2$ to $0.4$. Thus, we mark the progenitor systems using blue triangles based on a $\alpha_{\rm CE}$ value of 0.3 in this figure (except for ZTF~J0038+2030, for which a $\alpha_{\rm CE}$ value of 0.45 is used). Most of these triangles are located toward the longer end of the possible initial orbital period range. 
This is because a wider initial separation $a$ is required by the progenitor system when the energy releasing efficiency $\alpha_{\rm CE}$ is lower, which has been discussed in Sect.~\ref{sec:BP} when examining Figure~\ref{fig:m2_slice}.

It is clear from the plot that the two samples occupy different parameter spaces in the period-mass diagram.
The WD$-$BD progenitors derived in this work occupy mostly the parameter space with a period in the range of 50 to 600~days, while the known transiting or eclipsing MS$-$BD systems dominate the parameter space with a period in the range of $<50$~days. 
This is because most of these MS$-$BD sample are discovered in the transit exoplanet survey, which is biased toward detecting short-period systems. 
Thus, the new deduced WD$-$BD progenitor sample is a great complement to the directly observed transiting MS$-$BD sample used to study the shape of the BDD around A-F-G-type stars. 
After including the WD$-$BD progenitor binaries, the number of transiting or eclipsing BDs orbiting close to A-F-G-type stars with precise mass measurements has increased from 23 to 31, representing an approximate 34$\%$ increase in the sample size. 
\citet{Ma14} have found a low-density area in the period-mass diagram around 42.5~$M_{\rm Jupiter}$, which is proposed as an outcome of different formation mechanisms of low-mass BDs and high-mass BDs.
From Figure~\ref{fig:general}, we can see that this ``driest'' part in the BDD still appears to be underpopulated, extending into an orbital period of several hundred days. 
The difference between this study and previous studies is that we are using transiting or eclipsing BDs with precise masses measured instead of using BDs with only $M\sin i$ measured, thus making our results more robust.
However, it is still too early to give any final conclusion since this new derived sample is still small. 

In a recent study using {\it Gaia} Data Release 3 (DR3) data, \citet{Stevenson23} have identified 19 new BD candidates and updated BD masses for another 12 systems. 
These BD candidates exhibit similar locations to our reconstructed BDs in the period-mass diagram.
They find that BD companions with periods smaller than couple hundred days are still underpopulated in comparison with BDs with longer orbital periods, and there is a relatively flat mass valley around 25–45~$M_{\rm Jup}$.
Both studies show there is a lack of BDs with 10$-$40~$M_{\rm Jup}$ near orbital period around 50$-$300~d. This is likely caused by selection effects since astrometry is most sensitive to long-period orbits and massive objects, and it is easier to measure the mass of a more massive BD companion in a WD$-$BD binary. 
A better understanding of the detection biases of the WD$-$BD binaries can further improve our understanding of the shape of the BDD around A-F-G-type stars.
Thus, we urge our colleagues to present the study of survey detection biases in their future WD$-$BD discovery papers.

\subsection{Wide WD$-$BD binaries}
Besides the short-period post-CE WD$-$BD binaries selected in Section~\ref{sec:data}, wide and ultra-wide WD$-$BD binaries also have been discovered \citep{Becklin88, Farihi05, Steele09, Day-Jones11, Luhman11, Mace18, Zhang20, Wang23}. 
These binaries generally have a semimajor axis over $100$~AU, and were detected mainly from optical and infrared imaging survey and spectral survey.
We used the COMPAS code to study the possible formation channels for these wide WD$-$BD binaries. We find that the BD masses likely have not been altered through their lifetimes. 
However, according to our calculation, the orbital semimajor axis can become three times larger than the initial value due to mass ejection out of the binary system.
Thus, the progenitors of these wide or ultra-wide WD$-$BD binaries will also be wide or ultra-wide MS$-$BD binaries, and will not affect the shape of the desert in close orbits around MS A-F-G-type stars. 
Therefore, we decided to exclude these wide WD$-$BD binaries from this study. However, for research focusing on the distribution of wide BDs around MS stars \citep{Bowler20}, these wide WD$-$BD binaries should be included.

\section{Conclusion} \label{sec:conc}

Obtaining a large sample of BDs with precise mass measurements is important for constraining the shape of the BDD around solar-type stars \citep{Shahaf19}. 
Previous studies have focused on analyzing BD companions around MS stars.
In this work, we propose a new method for probing the shape of the BDD in the vicinity of MS A-F-G-type stars that uses post-CE WD$-$BD binaries. 
For each post-CE WD$-$BD system, we used a population synthesis technique to deduce the properties of its progenitor system, including the orbital period and mass of the primary star.
In this way, we added approximately ten objects with precise mass measurements to the period-mass diagram, which can be utilized to place additional constraints on the shape of the BDD around MS A-F-G-type stars. 
With many ongoing ground-based surveys, we expect the number of known post-CE WD$-$BD systems to increase rapidly in the near future. 

A distinctive aspect of this new sample is that most binaries exhibit orbital periods ranging from approximately 50 to 600~days, in contrast to the 1 to 50 days that are typical for the directly observed transiting MS$-$BD sample. 
This is because transit campaigns are usually strongly biased toward short-period systems. 
\citet{Stevenson23} used {\it Gaia} DR3 astrometry data to identify 19 BD candidates with periods of less than $\sim1200$~d, which is likely biased toward longer-period objects. 
Their orbital period distribution is more similar to our reconstructed systems. 
However, it is difficult to use the distribution of BDs from different surveys to study the location of the BDD due to observational biases and selection effects. 
Investigating the observational biases and selection effects of different surveys is beyond the scope of this paper. 
To facilitate such studies, we encourage our colleagues to report all relevant null detections in their future investigations.

\begin{acknowledgements}
      We acknowledge the financial support from the National Key R\&D Program of China (2020YFC2201400), NSFC grant 12073092, 12103097, 12103098, the science research grants from the China Manned Space Project (No. CMS-CSST-2021-B09). H.G. acknowledges support from the National Key R\&D Program of China (grant No. 2021YFA1600403), NSFC (grant No. 12288102, 12125303, 12173081), the key research program of frontier sciences, CAS, No. ZDBS-LY-7005, Yunnan Fundamental Research Projects (grant No. 202101AV070001).

\end{acknowledgements}

%


   \bibliographystyle{aasjournal}
   \bibliography{aanda} 

\end{document}